\documentclass[aps,pra,reprint,superscriptaddress,twocolumn]{revtex4}

\usepackage{graphicx,color}
\usepackage{epsfig}
\usepackage{amsmath, amssymb}
\usepackage{epstopdf}
\usepackage[T1]{fontenc}

\newcommand\UT{Department of Applied Physics, School of Engineering,\\ The University of Tokyo, 7-3-1 Hongo, Bunkyo-ku, Tokyo 113-8656, Japan}
\newcommand\UP{Department of Optics, Palack\'y University,
17. listopadu 1192/12, 77146 Olomouc, Czech Republic}
\newcommand\QPEC{Quantum-Phase Electronics Center, School of Engineering, The University of Tokyo,
7-3-1 Hongo, Bunkyo-ku, Tokyo 113-8656, Japan}

\begin{document}

\title{General implementation of arbitrary nonlinear quadrature phase gates}
\author{Petr Marek}
\email{marek@optics.upol.cz}
\affiliation{\UP}
\author{Radim Filip}
\affiliation{\UP}
\author{Hisashi Ogawa}
\affiliation{\UT}
\author{Atsushi Sakaguchi}
\affiliation{\UT}
\author{Shuntaro Takeda}
\affiliation{\UT}
\author{Jun-ichi Yoshikawa}
\affiliation{\UT}
\affiliation{\QPEC}
\author{Akira Furusawa}
\email{akiraf@ap.t.u-tokyo.ac.jp}
\affiliation{\UT}

\date{\today}

\begin{abstract}
We propose general methodology of deterministic single-mode quantum interaction nonlinearly modifying single quadrature variable of a continuous variable system. The methodology is based on linear  coupling of the system to ancillary systems subsequently measured by quadrature detectors. The nonlinear interaction is obtained by using the data from the quadrature detection for dynamical manipulation of the coupling parameters.
This measurement-induced methodology enables direct realization of arbitrary nonlinear quadrature interactions
without the need to construct them from the lowest-order gates. Such nonlinear interactions are crucial for more practical and efficient manipulation of continuous quadrature variables as well as qubits encoded in continuous variable systems.
\end{abstract}

\maketitle


Quantum technology employing quantum information processing with qubits is constrained to potentially large but always finite-dimensional Hilbert spaces \cite{DVprocessing}. To move beyond this limitation and fully process and simulate infinite dimensional systems one has to take advantage of continuous variables (CV) methods \cite{CVprocessing, CVsimulation}.
Moreover, CV methods are suitable for manipulating qubits encoded in the subspace of infinite dimensional systems \cite{GKP,hybrid}.
Such a hybrid qubit-CV approach has turned out to have practical advantages in quantum optics
since it can take advantage of robust encoding of qubits and deterministic operation with CV methods \cite{hybrid,HybridTele}.
The experimentally accessible CV operations are linear transformations of continuous quadrature operators and can be constructed from Hamiltonians of up to quadratic order of the operators \cite{LinearProcessing}.
Such linear transformations can be deterministically performed for systems in both Gaussian and non-Gaussian states \cite{SPsqueezing}.
They cannot, however, provide the nonlinear non-Gaussian dynamics which is necessary for accessing the full quantum analog simulation \cite{CVsimulation} and computation \cite{CVprocessing}. For that we require elementary nonlinear transformations which require Hamiltonians with cubic or higher order nonlinearity \cite{Lloyd}.


Gottesman, Kitaev, and Preskill (GKP) stimulated long-standing theoretical and experimental development of the missing tools required for the elementary third order (cubic) nonlinear phase gate \cite{GKP}.
We have recently expanded upon the original concept by designing a deterministic cubic nonlinear phase gate for a traveling beam of light based on adaptive continuous-variable measurement and linear feed-forward control \cite{MarekX3}.
In principle, this nonlinear cubic gate, together with already existing linear and quadratic gates, is sufficient for constructing an arbitrary nonlinear gate and realizing universal computing with CVs \cite{Lloyd}.
This gate set also enables deterministic and universal quantum computation for qubits with the hybrid approach \cite{hybrid}.
However,  useful gates for qubits and CVs are often of higher order and require impractical number of elementary gates for implementation \cite{SefiCommute2, SefiCommute3}.
For example, the fourth order Kerr nonlinearity is necessary for realizing
controlled-NOT gates of qubits, quantum nondemolition measurement of photon number \cite{QND}, and creation of Schr\"{o}dinger cat states \cite{CatGeneration},
but its implementation with sufficiently small errors requires tens of individual cubic or lower order gates \cite{SefiCommute2}.
As the order of the desired nonlinearity increases, the number of required gates rapidly increases and soon becomes experimentally intractable.

In this letter we present a full methodology for
directly realizing deterministic nonlinear quadrature phase gates of an arbitrary order.
These gates require a set of ancillary harmonic oscillators linearly coupled to the target system and measured by quadrature detectors. The required nonlinearity is obtained by nonlinear classical feed-forward control \cite{dynamicFF}.
In order to compensate quantum noise appearing due to the deterministic nature of the gates, the ancillary oscillators need to be initialized in \emph{nonlinearly squeezed states}. Such states can be prepared in advance by probabilistic methods \cite{NonlinearStates} or on different platforms, and stored before they are needed \cite{Memory,synchronization}. We will describe the overall strategy and then focus on the illustrative example of the fourth order (quartic) nonlinear gate.
The proposal is implementable with the current optical hybrid technology \cite{hybrid}, making it suitable for efficient realization of universal quantum computing with qubits and CVs.
It can be also adapted to other physical platforms, such as phononic modes in quantum electromechanical and optomechanical systems \cite{Oscillators}, motion modes of trapped ions \cite{Ions}, microwave radiation in cavity QED \cite{Microwaves}, or collective spins of atoms \cite{AtomicSpins, Opatrny}.


\emph{CV quantum operations} -
The ultimate tool of CV quantum information processing is a unitary transformation realizing dynamics of an arbitrary Hamiltonian \cite{Lloyd}. For CV harmonic oscillators, which are described with help of quadrature operators $\hat{x}$ and $\hat{p}$, with $[\hat{x},\hat{p}] = i$, the arbitrary Hamiltonian can be expressed as a bivariate polynomial $\hat{H} = \sum_{k,l} c_{k,l} (\hat{x}^k \hat{p}^l + \hat{p}^l \hat{x}^k)$. The elementary technique that allows construction of such operators relies on using a number of simple operations and merging them together as:
\begin{equation}\label{commutation}
    e^{iA}e^{iB}e^{-iA}e^{-iB} \approx e^{\frac{i}{2}[A,B]}.
\end{equation}
This technique, originally presented in \cite{Lloyd} and in larger detail studied in \cite{SefiCommute2,SefiCommute3}, allows combining operations with different Hamiltonians into their composites. When the orders of the constituent Hamiltonians are $N_A$ and $N_B$, the resulting Hamiltonian is of the order $N = N_A + N_B - 2$. This means that combining operations of at least third  order is capable of creating an operation with order higher than that of its constituents, which can ultimately lead to creation of operations with arbitrary orders. The most elementary operation suitable for this operation is the cubic phase gate with Hamiltonian $\hat{H} \propto \hat{x}^3$ \cite{Lloyd, MarekX3}. However, as the order of the desired operation grows, we can start encountering scaling issues. The exact quantity of required operations strongly depends on their specific forms, but, for example, realizing operation of 10th order requires \textit{at least} $2^6$ individual third order operations \cite{Explanation1}.
This issue could be resolved by realizing at least some of the higher order operations directly, without the need to construct them from the lowest level components repeatedly using formula (\ref{commutation}). 
\begin{figure}
\begin{center}
\includegraphics[width=0.8\linewidth]{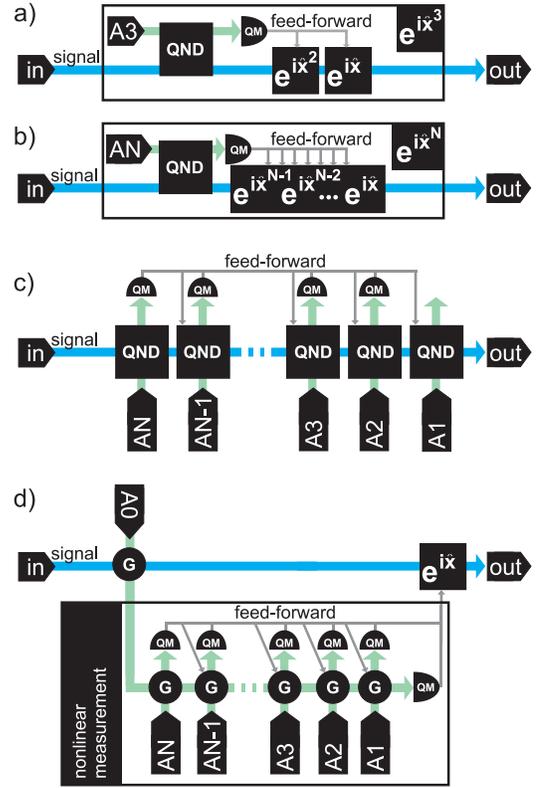}
\end{center}
\caption{Schematic circuits for various implementations of nonlinear gates. QND - quantum non-demolition interaction, QM - quadrature measurement, $Ak$ - ancillary state of the $k$-th order squeezed in $\hat{p} - N\chi_N \hat{x}^{N-1}$.
$e^{i\hat{x}^k}$ -
unitary realization of $k$-th order nonlinear gate with arbitrary strength. a) Cubic gate with $N = 3$; b) $N+1$-th order gate implemented recursively; c) $N$-th order gate with streamlined feed-forward; d) $N$-th order gate implemented in the measurement induced way. $G$ represents a tunable Gaussian operation, which can be either QND or beam splitter. $A0$ is ancillary state squeezed in $\hat{x}$.
}\label{fig_setups}
\end{figure}

In the Heisenberg representation, the cubic phase gate transforms operators of a quantum state as  $\hat{x}'=\hat{x}$ and
$\hat{p}'=\hat{p}-3\chi_3 \hat{x}^2$, where $\chi_3$ is the cubic interaction gain. The realizing quantum circuit is depicted in Fig.~\ref{fig_setups}a.
The two oscillators, the signal and the ancilla, are coupled through a QND gate, which is characterized by interaction Hamiltonian $H_{\mathrm{QND}} = \hat{x} \hat{p}_a$. The $\hat{x}_a$ quadrature of the ancilla is then measured and the obtained value is used to drive feed-forward corrections of the first (displacement) and second (squeezing) orders. The coupling and the feed-forward operations are individually Gaussian, but the ancillary state $A3$ is not. In order to compensate for the back action noise, the ancilla $A3$ has to be prepared in the \emph{cubic squeezed state}, which has fluctuations of operator
$\hat{p}_a - 3\chi_3 \hat{x}_a^2$, where the parameter $\chi_3$ sets the strength of the nonlinearity, below the vacuum level,
$\langle ([\Delta(\hat{p}_a-3\chi_3\hat{x}^2_a)]^2 \rangle<0.5$, and ideally approaching zero.


The principle can be extended to  nonlinear Hamiltonians of higher order, $\hat{H} \propto \hat{x}^N$. They can be realized by employing an ancilla with reduced fluctuations in quadrature $\hat{p} -
N\chi_N \hat{x}^{N-1}$. However, in this case, the required feed-forward operations are of orders $1,\cdots,N-1$, see Fig.~\ref{fig_setups}b and each of them requires an  ancilla squeezed in a specific nonlinear quadrature. So, while the same method can be used for realizing these lower-order nonlinear circuits in such recursive manner, the total number of gates required for realizing operation of $N$-th order is $2^{N-3}$, which is again the undesirable exponential scaling.

Fortunately it is possible to merge the required feed-forward operations so that only $N-2$ individual nonlinear gates are needed in total.
The scheme is depicted in Fig.~\ref{fig_setups}c and it relies on a sequence of $N$ QND interactions with $N$ ancillary states with reduced fluctuations in quadratures $\hat{p}_{Ak} - \hat{x}_{Ak}^{k-1}$, where $k = 1,\cdots,N$. It is a significant advantage that the gains of the Gaussian QND operations depend on the previous results while the states do not. Setting the proper QND gains can be realized by fast feed-forward \cite{dynamicFF}, which is significantly more feasible than preparing tailored quantum states.
Also, for $k= 1,2$ the required states are Gaussian and the gates are not nonlinear. As a consequence, the required operation can be usually realized in a different manner \cite{adaptive_measurement}. For the sake of resulting formulas, though, we are going to use the gate-based expression. The QND operations transform the quadrature operators of the signal $s$ and the $k$-th ancillary mode $Ak$ according to
\begin{eqnarray}\label{}
    \hat{x}'_s = \hat{x}_s, \hat{p}'_s = \hat{p}_s + z_k \hat{p}_{Ak},\nonumber \\
    \hat{x}'_{Ak} = \hat{x}_{Ak} - z_k \hat{x}_s, \hat{p}'_{Ak} = \hat{p}_{Ak}.
\end{eqnarray}
The ancillary modes are then measured, yielding values $q_k = \hat{x}_{Ak} - z_k \hat{x}_s$. The gains $z_k$ of the QND operations are going to be functions of the previously measured values. To find them, we can express the final quadrature relations as
\begin{equation}\label{relation1}
    \hat{x}_{out} = \hat{x}_{in}, \quad\hat{p}_{out} = \hat{p}_{in} + \sum_{j = 0}^{N-1} z_j \hat{p}_{Aj},
\end{equation}
where $z_j$ are yet to be determined. We can use the nonlinear property of ancillary states and the relationship between the operators and the measured quadratures,
\begin{equation}\label{}
    \hat{p}_{Ak} = \hat{x}_{Ak}^{k-1},\quad \hat{x}_{Ak} = z_k \hat{x}_{in} + q_k,
\end{equation}
where $q_k$ are the values obtained by the quadrature detectors, and arrive at the final form of the $\hat{p}$-quadrature relations as
\begin{eqnarray}\label{relation2}
\hat{p}_{out} &=&  \hat{p}_{in} + \sum_{k = 0}^{N-1} \hat{x}_{in}^k \sum_{j = 1}^{N-k}
                                                                                        \nonumber \\
& \times & \left(\begin{array}{c}
                                                                                            N-j \\
                                                                                            k
                                                                                          \end{array}\right)
                                                                                          (q_{N-j+1})^{N-j-k} (z_{N-j+1})^{k+1}.
\end{eqnarray}
We can see that transformation given by (\ref{relation1}) and (\ref{relation2}) realizes the desired $\hat{x}^N$ operation when the QND gain is proportional to the desired nonlinear operation gain,  $(z_{N})^N = \chi_{N}$, and the remaining gains satisfy a set of $N-1$ equations
\begin{equation}\label{}
    \sum_{j = 1}^{N-k} \left(\begin{array}{c}
     N-j \\
       k
      \end{array}\right)
     (q_{N-j+1})^{N-j-k} (z_{N-j+1})^{k+1} = 0
\end{equation}
for all $k = 0,\cdots,N-2$. This is a set of polynomial equations for $z_j$ which is already in the upper diagonal form and has always a unique solution. More importantly, the solution can be found in a recurrent form, so value of each $z_j$ is function only of the already known quantities $z_m$ and $q_m$, where $m>j$. Also note that the measured value $q_1$ is not needed and the measurement therefore does not need to be performed. This shows that the GKP approach can be extended for realization of an arbitrary order of the $\hat{x}^N$ gate and that the extension can be performed in such the way to efficiently resolve the scaling issues. 

\emph{Nonlinear measurement induced approach - }
Applying elementary quantum circuits directly to a quantum state is a very straightforward approach. However, in practice it is often beneficial to take advantage of the inherent entangling property of quantum states and impress the desired nonlinearity onto the states through a suitable measurement performed on a suitable subsystem. So while the components of the circuit in Fig.~\ref{fig_setups}c already follow the measurement induced paradigm, it is sensible to take this path to its logical conclusion and perform the full gate completely through a measurement. The scheme is sketched in Fig.~\ref{fig_setups}d and it consists of a single QND interaction coupling together the initial system with ancillary system $As$ prepared in a sufficiently squeezed vacuum state. This ancillary system is then subjected to the in-line non-linear gate consisting of QND gates with parameters $z_k$ coupling the system to $N$ ancillary states, which are subsequently measured by $\hat{x}$-quadrature detectors. In addition, the remaining ancillary mode is measured by a $\hat{p}$-quadrature measurement, which is used to erase the influence of the carrier ancilla. The individual $\hat{x}$-quadrature measurements provide measurement results $q_k = \hat{x}_k - z_k \hat{x}_{in}$. After the initial system is displaced by the measured value of the final $\hat{p}$-quadrature measurement, $y = \hat{p}_0 + \sum_{k = 1}^{N}\hat{p}_{Ak}$, the quadrature operators of the initial system can be exactly described by (\ref{relation1}) and therefore subsequently corrected in the same manner. Under ideal conditions the measurement induced and the in-line schemes are mathematically equivalent.

The QND coupling can be also replaced by a symmetric passive linear coupling, which is described by interaction Hamiltonian $\hat{H}_{BS} \propto \hat{x}_1\hat{p}_2 + \hat{p}_1\hat{x}_2$. This coupling, which for optical systems stands for the ubiquitous beam splitter, is passive; it only transfers energy between the systems instead of creating it. As a consequence it often is more feasible and less prone to noise and imperfections, and at optical frequencies it can work with arbitrarily high speed. On the other hand, the mixing of both quadratures makes it often more difficult to treat, as compared to the QND. In our scenario, however, the operations can be made equivalent. To see this, let us again consider the measurement induced scheme of Fig.~\ref{fig_setups}d. The first beam splitter can have an arbitrary transmissivity $t_0$. However, we will also it preceded by Gaussian squeezing operation, which ensures that $\hat{x}_{out} = \hat{x}_{in}$. After the ancillary state $A0$ interacts with the first beam splitter, with positive transmissivity $t_{N}$ and reflectivity $r_{N}$, it transforms to
\begin{equation}\label{state_continues}
    \hat{x}_s^{(N)} = t_{N}\hat{x}_{in} + r_{N}\hat{x}_{AN},\quad \hat{p}_s^{(N)} = t_{N}\hat{p}_{in} + r_{N}\hat{p}_{AN},
\end{equation}
and the $\hat{x}$ quadrature measurement of the nonlinear ancilla provides value $q_{N} = t_{N}\hat{x}_{AN} - r_{N}\hat{x}_{in}$. In order to simplify the description we can now use this measured value and use it to transform the state (\ref{state_continues}) by Gaussian displacement and squeezing into:
\begin{equation}\label{connectlink}
     \hat{x}_s^{(N)'} = \hat{x}_{in},\quad \hat{p}_s^{(N)'} = t_{N}^2\hat{p}_{in} + t_{N}r_{N}\hat{p}_{AN}.
\end{equation}
Since these operations are Gaussian, as is the rest of the active components of the circuit, it is enough to consider them virtually and include their influence only into the measured data. Here we treat them as physical operations to simplify the derivation. After the sequence of all $N$ beam splitters and erasing the influence of the carrier ancilla, the quadrature operators of the signal can be expressed as
\begin{eqnarray}\label{}
    \hat{x}_{out} = \hat{x}_{in},\quad \hat{p}_{out} = \hat{p}_{in} + \sum_{j=1}^{N}\left(t_j r_j \prod_{k = j}^{N}t_k^{-2}\right) \hat{p}_{Aj}.
\end{eqnarray}
The form is again equivalent to (\ref{relation1}). The coefficients $t_j r_j \prod_{k = j}^{N}t_k^{-2}$ which need to be compensated are more involved than in the previous scenarios, but the final set of equations for the beam splitter coefficients can be solved in the same manner as for the QND scenario.

\begin{figure}
\centering
\includegraphics[width=0.5\linewidth]{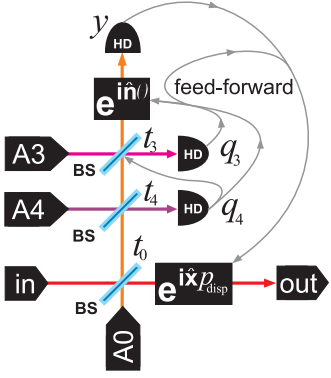}
\caption{(Color online) Scheme for the optical realization of the fourth order nonlinear circuit. BS - beam splitter; HD - homodyne detection; $e^{i\hat{n}\theta}$ - operation realizing $\theta$ phase shift; $e^{i\hat{x}p_{\mathrm{disp}}}$ - $\hat{p}$-quadrature displacement by value $p_{\text disp}$; $t_0$, $t_4$, $t_3$ - splitting ratios of respective beam splitters; $y$, $q_4$, $q_3$ - values measured by the homodyne detectors.  }
\label{schemeX4}
\end{figure}

\emph{Quartic nonlinearity.}
This specific gate, a step above the elementary cubic nonlinearity, is strongly beneficial in realization of Kerr nonlinearity \cite{SefiCommute3}.
The particular linear optical scheme is in Fig.~\ref{schemeX4}.
The implementation follows the steps drawn in the general section with only few differences. The ancillary states are prepared with squeezing in quadratures $\hat p_{Ak}- k \chi_k\hat x^{k-1}_{Ak}$, where the parameters $\chi_k$ are not related to the strength of the nonlinearity and only represent additional degrees of freedom which can be exploited during the preparation. The squeezing operations (\ref{connectlink}) previously considered to simplify the description are missing. The last two blocks corresponding to ancillas of orders $1$ and $2$ are also missing; these two operations are Gaussian and are therefore implemented in another way. The displacement directly, the squeezing by adaptive measurement of the quadrature rotated by $\theta$, which depends on previous measurement results \cite{adaptive_measurement}. The three values measured by the optical homodyne detectors are:
\begin{align}\label{measured_values}
  q_4 &= -r_0r_4\hat x_\text s
     -t_0r_4 \hat x_{A0}
     + t_4\hat x_{A4}, \\
  q_3 &= -r_0t_4r_3\hat x_\text s
     -t_0t_4r_3 \hat x_{A0}
     -r_4r_3\hat x_{A4}
     +t_3\hat x_{A3}, \\
    y &=\sin\theta (r_0t_4t_3\hat x_\text s
     +t_0t_4t_3 \hat x_{A0}
     +r_4t_3\hat x_{A4}
     +r_3\hat x_{A3}) \nonumber \\
&+\cos\theta (r_0t_4t_3\hat p_\text s
     +t_0t_4t_3 \hat p_{A0}
     +r_4t_3\hat p_{A4}
     +r_3\hat p_{A3}).
\end{align}
The splitting ratio of the second beam splitter, as well as the required phase shift, depend on the already measured results:
\begin{eqnarray}\label{}
    \chi_3\left(\frac{r_3}{t_3}\right)^3=-\frac{4\chi_4r_4^3}{t_4}q_4, \\
    \tan\theta = -\frac{6\chi_3r_3^2}{t_3}q_3-\frac{12\chi_4 r_4^2t_3^2}{t_4^2}\left(t_4^2 -r_4^2\right)q_4^2,
\end{eqnarray}
and fast electronic circuits \cite{hybrid} are required to process the data quickly enough to provide the required feed-forward. Finally, the remaining signal state needs to be displaced by a single value
\begin{align}
p_{\text{disp}} &= -\frac{4\chi_4r_0r_4}{t_0t_4^4}q_4^3-\frac{3\chi_3r_0r_3}{t_0t_4t_3^3}\left(\frac{r_4r_3}{t_4}q_4+q_3\right)^2\nonumber\\
&-\frac{r_0r_4}{t_0t_4^2t_3^2}\tan \theta \left(q_4+\frac{t_4r_3}{r_4}q_3\right)+\frac{r_0}{t_0t_4t_3\cos\theta}q_2
\end{align}
in order to transform the output quadrature operators to
\begin{subequations}
\label{eq:x4withNoise}
\begin{align}
\hat x_{\text out}&= t_0 \hat x_{\text in}- r_0 \hat x_{A0}, \label{eq:x4withNoiseX}\\
\hat p_{\text out} &=\frac{1}{t_0}\left[\hat p_{\text in} + \frac{4\chi_4r_0^4r_4^4}{t_4^4}\left(\hat x_{\text in}+\frac{t_0}{r_0}\hat x_{A0}\right)^3\right]\nonumber\\
&+\frac{r_0r_4}{t_0t_4}\left(\hat p_{A4}-4\chi_4\hat x^3_{A4}\right)\nonumber\\
&+\frac{r_0r_4}{t_0t_4}\left[\frac{4\chi_4}{\chi_3t_4}\left(r_0r_4\hat x_\text s+t_0r_4 \hat x_{A0}-t_4\hat x_{A4}\right)\right]^{\frac{1}{3}}\nonumber\\
&\times\left(\hat p_{A3}-3\chi_3\hat x^2_{A3}\right).
\label{eq:x4withNoiseP}
\end{align}
\end{subequations}
We can see that the operators correspond to the input signal, squeezed by factor $t_0$, transformed by the fourth order nonlinear phase gate with effective strength $\chi_4' = \frac{4\chi_4r_0^4r_4^4}{t_4^4}$. The remaining terms represent the imperfections arising from ancillary states - both the finite linear squeezing in the mode $A0$ and the finite nonlinear squeezing in modes $A4$ and $A3$. The last term depends on both nonlinear ancillas, which is caused by coupling parameter $t_3$ depending on the measurement of $A4$.  As a consequence, for good performance the nonlinear ancillary states should satisfy
\begin{align}
\langle[\Delta(\hat p_{A3}-3\chi_3\hat x^2_{A3})]^2\rangle\ll \frac{1}{\langle [\Delta \hat x_{A4}^\frac{1}{3}]^2\rangle}.
\end{align}
This represents an example of squeezing requirement for a new class of nonlinear squeezed states. The dynamical problem of implementing any nonlinear phase gate has been therefore turned into the static problem of preparing suitable quantum resource states.

\emph{Conclusion.}
The presented methodology has two revolutionary advantages over the previous methods. First, further integration of feed-forward to adjust the coupling coefficients allows to manipulate with strengths of the nonlinear operation by using only Gaussian tools. As a consequence, there is no need to prepare nonlinear quantum states for specific strengths of the nonlinearity, which significantly streamlines the state preparation phase of the circuit, as it moves all non-Gaussian requirements to preparation of \emph{only} universal single-mode nonlinear squeezed states.
Second, the ability to merge the necessary feed-forwards into a single sequence removes the exponential scaling in the number of operations.
Together these innovations with the current development of time-resolved optical quantum technology \cite{synchronization} open up the possibility of feasible and efficient experimental realization of the nonlinear phase gates and their application to CV simulation and hybrid qubit-CV computation \cite{CVsimulation,CVprocessing,hybrid}.

\section*{Acknowledgement}
This work was partly supported by CREST (JPMJCR15N5) of JST, JSPS KAKENHI, APSA.
H. O. acknowledges financial support from ALPS. P. M. and R. F. acknowledge Project GB14-36681G of the Czech Science Foundation.


\begin{thebibliography}{cc}
\bibitem{Explanation1}
Operation of 10th order requires $2^2$ 6th order operations, which is $2^4$ 4th order operations or $2^6$ 3rd order operations.
\end{thebibliography}
\end{document}